\documentclass[11pt]{iopart2}

\usepackage{graphicx,bm,float,amsmath,amssymb}

\def\spose#1{\hbox to 0pt{#1\hss}}
\def\lesssim{\mathrel{\spose{\lower 3pt\hbox{$\mathchar"218$}}
 \raise 2.0pt\hbox{$\mathchar"13C$}}}
\def\gtrsim{\mathrel{\spose{\lower 3pt\hbox{$\mathchar"218$}}
 \raise 2.0pt\hbox{$\mathchar"13E$}}}

\def\<{\langle}
\def\>{\rangle}

\usepackage[english]{babel}
\usepackage[T1]{fontenc}
\usepackage[utf8]{inputenc}

\begin{document}

\title{Large-$N$ behavior of three-dimensional lattice CP$^{N-1}$ models}

\author{Andrea Pelissetto$^1$ and Ettore Vicari$^2$}
\address{$^1$ Dipartimento di Fisica dell'Universit\`a di Roma Sapienza
        and INFN Sezione di Roma I, I-00185 Roma, Italy}
\address{$^2$ Dipartimento di Fisica dell'Universit\`a di Pisa
        and INFN Largo Pontecorvo 3, I-56127 Pisa, Italy}

\ead{Andrea.Pelissetto@roma1.infn.it,Ettore.Vicari@unipi.it}

\begin{abstract}
We investigate the phase diagram and critical behavior of a
three-dimensional lattice CP$^{N-1}$ model in the large-$N$ limit.
Numerical evidence of first-order transitions is always observed
for sufficiently large values of $N$, i.e. $N>2$ up to $N=100$.
The transition becomes stronger---both the latent heat and the surface
tension increase---as $N$ increases. Moreover, on the high-temperature
side, gauge fields decorrelate on distances of the order of one
lattice spacing for all values of $N$ considered. Our results are
consistent with a simple scenario, in which the transition is of first
order for any $N$, including $N\to\infty$. We critically discuss the
analytic large-$N$ calculations that predicted a large-$N$ continuous
transition, showing that one crucial assumption made in these
computations fails for the model we consider.
\end{abstract}

\maketitle


\section{Introduction}
\label{intro}

Models that are invariant under a local U(1) gauge symmetry have been
systematically studied in condensed-matter and in high-energy physics.
The simplest model with a local U(1) gauge symmetry is the CP$^{N-1}$
model.  In three dimensions it emerges as an effective theory
describing several condensed-matter
systems~\cite{RS-90,TIM-05,Kaul-12,KS-12,BMK-13,MV-04,SBSVF-04}, while
in two dimensions it is an interesting theoretical laboratory to study
quantum field theories of fundamental interactions as it shares
several features with quantum chromodynamics (QCD), the theory that
describes the hadronic strong interactions~\cite{ZJ-book,MZ-03}.

A lattice formulation of the CP$^{N-1}$ model is obtained by
associating complex $N$-component unit vectors ${\bm z}_{\bm x}$ with
the sites ${\bm x}$ of a cubic lattice, and U(1) variables
$\lambda_{{\bm x},\mu}$ with each link connecting the site ${\bm x}$
with the site ${\bm x}+\hat\mu$ (where
$\hat\mu=\hat{1},\hat{2},\ldots$ are unit vectors along the lattice
directions).  The partition function of the system reads
\begin{equation}
Z = \sum_{\{{\bm z}\},\{\lambda\}} e^{-\beta H}\,,
\label{partfun}
\end{equation}
where the Hamiltonian is
\begin{equation}
H = - N \sum_{{\bm x}, \mu} \left( \lambda_{{\bm x},\mu}\,
\bar{\bm{z}}_{\bm x} \cdot {\bm z}_{{\bm x}+\hat\mu} + {\rm
  c.c.}\right)\,,
\label{Ham}
\end{equation}
and the sum runs over all lattice links.  Such a model can also be
seen as a limiting case of the lattice abelian Higgs model in which
the gauge fields become dynamical, see \cite{PV-19-AH} and references
therein.

In three dimensions CP$^{N-1}$ models are expected to undergo a
finite-temperature transition. The associated order parameter is the
gauge-invariant quantity
\begin{equation}
Q_{\bm x}^{ab} = \bar{z}_{\bm x}^a z_{\bm x}^b - {1\over N} \delta^{ab} \,.
\label{Q-def}
\end{equation}
In the high-temperature (HT) phase, the system is disordered and
$\langle Q^{ab}\rangle = 0$, while in the low-temperature (LT) phase
the parameter $Q^{ab}$ magnetizes.  In spite of extensive
field-theoretical and numerical studies for $N=2,3,4$ and
$N\to\infty$, the nature of the transition is still controversial
\cite{MZ-03,NCSOS-11,NCSOS-13,PV-19}, and in particular, it is not
clear whether CP$^{N-1}$ universality classes exist for $N>2$.

Model (\ref{Ham}) was numerically studied in \cite{PV-19} for
$N=2,3,4$.  It was shown that the system undergoes a continuous
transition for $N=2$, in the O(3) universality class. For
$N=3$ and 4, it undergoes a first-order transition. No numerical
results are available for larger values of $N$, although both existing
lattice and continuum analytic computations predict a continuous
transition for $N=\infty$. Consistency between the large-$N$
prediction and the $N=3,4$ numerical results requires the existence of
a critical number of components $N_c$, such that the transition is of
first order for $2<N< N_c$ and continuous for $N>N_c$. The universality
class of the critical transition for $N>N_c$ would then be naturally
identified with that associated with the large-$N$ fixed point
occurring in the continuum abelian-Higgs model
\cite{HLM-74,FH-96,IZMHS-19} and which exists for $N \ge N_{c0}$. 
The estimate $N_{c0} = 12.2(3.9)$ was obtained by an analysis of the
  $\epsilon$ expansion up to four loops~\cite{IZMHS-19}.  Of course,
the critical number $N_c$ should be larger than or equal to
$N_{c0}$. Therefore, a numerical study of the model for quite large
values of $N$ is required.

The purpose of this paper is that of determining the nature of the
transition in the model with Hamiltonian (\ref{Ham}) for large values
of $N$.  We report results for $7\le N \le 100$, which are all
consistent with a simple scenario in which the transition is always of
first order, even for $N\to\infty$. This conclusion contradicts the
analytic calculations for $N=\infty$, forcing us to review the 
assumptions that are generally made in the standard large-$N$ approach
\cite{MZ-03,CR-93,KS-08,PV-19}. In particular, we verify that one crucial assumption
in these calculations is not correct. All calculations assume that the
gauge fields order for $N\to \infty$, i.e., that one can set
$\lambda_{{\bm x},\mu} = 1$ in this limit. We find that this
assumption is correct in the low-temperature phase, but not in the
high-temperature phase.  In the latter one, gauge fields as well as
gauge-invariant observables remain uncorrelated up to the transition
point. The transition is therefore of first order.

The paper is organized as follows. In Sec.~\ref{sec2} we present our
numerical large-$N$ results. In Sec.~\ref{sec2.1} we give some details
on the numerical simulations and define the observables we measure in
the Monte Carlo (MC) simulations. In Sec.~\ref{sec2.2} we present the
numerical results at the transition, while in Sec.~\ref{sec2.3} we
discuss the nature of the two phases. In Sec.~\ref{largeN-sol} we
review the large-$N$ analytic calculations. Finally, in
Sec.~\ref{Conclusions} we summarize and present our conclusions.

\section{Numerical results}
\label{sec2}

\subsection{Numerical simulations and observables}
\label{sec2.1}

In this section we present numerical results for systems with $7\le N
\le 100$.  We perform MC simulations on cubic lattices
of linear size $L$ with periodic boundary conditions, using the same
overrelaxed algorithm as in our previous work \cite{PV-19}.

We compute the energy density and the specific heat, defined as
\begin{equation}
E = {1\over N V} \langle H \rangle\,,\qquad
C ={1\over N^2 V}
\left( \langle H^2 \rangle 
- \langle H  \rangle^2\right)\,,
\label{ecvdef}
\end{equation}
where $V=L^3$.  We consider correlations of the hermitean gauge
invariant operator (\ref{Q-def}).  Its two-point correlation function
is defined as
\begin{equation}
G({\bm x}-{\bm y}) = \langle {\rm Tr}\, Q_{\bm x} Q_{\bm y} \rangle\,,  
\label{gxyp}
\end{equation}
where the translation invariance of the system has been taken into
account.  The susceptibility and the correlation length are defined as
$\chi=\sum_{\bm x} G({\bm x})$ and
\begin{equation}
\xi^2 \equiv  {1\over 4 \sin^2 (\pi/L)}
{\widetilde{G}({\bm 0}) - \widetilde{G}({\bm p}_m)\over 
\widetilde{G}({\bm p}_m)}\,,
\label{xidefpb}
\end{equation}
where $\widetilde{G}({\bm p})=\sum_{{\bm x}} e^{i{\bm p}\cdot {\bm x}}
G({\bm x})$ is the Fourier transform of $G({\bm x})$, and ${\bm p}_m =
(2\pi/L,0,0)$. We also consider the Binder parameter
\begin{equation}
U = {\langle \mu_2^2\rangle \over \langle \mu_2 \rangle^2} \,, \qquad
\mu_2 = 
\sum_{{\bm x},{\bm y}} {\rm Tr}\,Q_{\bm x} Q_{\bm y}\,,
\label{binderdef}
\end{equation}
and vector correlations of the fundamental variable ${\bm z}_x$.  We
define
\begin{equation}
G_V(\ell,L) ={1\over V} 
   \sum_{{\bm x}} 
  \hbox{Re}\left\langle
  \bar{\bm z}_{\bm x}\cdot {\bm z}_{{\bm x}+\ell \hat{\mu}}  
   \prod_{n=0}^{\ell-1} \lambda_{{\bm x}+n \hat{\mu},\mu}
   \right\rangle,
   \label{Gd}
\end{equation}
where all coordinates should be taken modulo $L$ because of the
periodic boundary conditions.  Note that in the definition (\ref{Gd})
we average over all lattice sites ${\bm x}$ exploiting the translation
invariance of systems with periodic boundary conditions, and select a
generic lattice direction $\hat{\mu}$ (in our MC simulations we also
average over the three equivalent directions).  Note also that
$G_V(0,L) = 1$ and that $G_V(L,L)$ is the average value $P(L)$ of the
Polyakov loop,
\begin{equation}
   P(L) = {1\over V} \sum_{\bm x} 
   \hbox{Re } \left\langle \prod_{n=0}^{L-1} \lambda_{{\bm x}+n \hat{\mu},\mu}
   \right\rangle\,.
   \label{Polyakov}
\end{equation}

\subsection{Behavior at the transition point}
\label{sec2.2}

In our previous work \cite{PV-19} we showed that the CP$^1$ model with
Hamiltonian (\ref{Ham}) undergoes a continuous transition, while the
CP$^2$ and CP$^3$ models have a first-order transition. We now
consider $N=7,10,15$, and 20. We find that the transition is of first
order in all cases, with a latent heat that increases with increasing
$N$: the transition becomes stronger as $N$ increases.

To determine the position of the transition and ascertain its order,
we use the phenomenological theory presented in \cite{CLB-86,VRSB-93}.
If the transition is of first order, at fixed $L$ the specific heat
$C$ and the Binder parameter $U$ have maxima $C_{\rm max}(L)$ and
$U_{\rm max}(L)$, respectively, which are proportional to the system
volume $V$. For $L\to \infty$, we have
\begin{equation}
C_{\rm max}(L) =V\left[ {1\over 4} \Delta_h^2 + O(V^{-1})\right]\,,\qquad
U_{\rm max}(L) = V\left[ a + O(V^{-1})\right]\,,
\label{CmaxUmax}
\end{equation}
here $\Delta_h$ is the latent heat, defined as $\Delta_h =
E(\beta\to\beta_c^+) - E(\beta\to\beta_c^-)$.  The values $\beta_{{\rm
    max},C}(L)$ and $\beta_{{\rm max},U}(L)$ where the maximum is
attained converge to the transition inverse temperature $\beta_c$ as
\begin{equation}
\beta_{{\rm max},C}(L)-\beta_c\approx c_1\,V^{-1}\,,\qquad \beta_{{\rm
    max},U}(L)-\beta_c\approx c_2\,V^{-1}\,.
\end{equation}
For each value of $N$ we determine the temperatures at which the
specific heat $C$ and the Binder parameter $U$ have a peak, and then we
study the behavior of the maxima as a function of $V$, to infer
the order of the transition. In the presence of a first-order
transition one should carefully verify that the simulation correctly
samples both phases. As we are using a local Metropolis/microcanonical
algorithm, this only occurs if the barrier between the two phases is
not too high; otherwise, the system is trapped in the phase in which
the simulation is started. Since, as we shall discuss, the first-order
transition is strong for the values of $N$ we consider, we have been
limited to relatively small lattices. In practice, we have results for
$L\le 12$, 10, 8, 6, for $N=7$, 10, 15, and 20.

\begin{figure}[tbp]
\begin{center}
\includegraphics[width=0.5\textwidth, keepaspectratio]{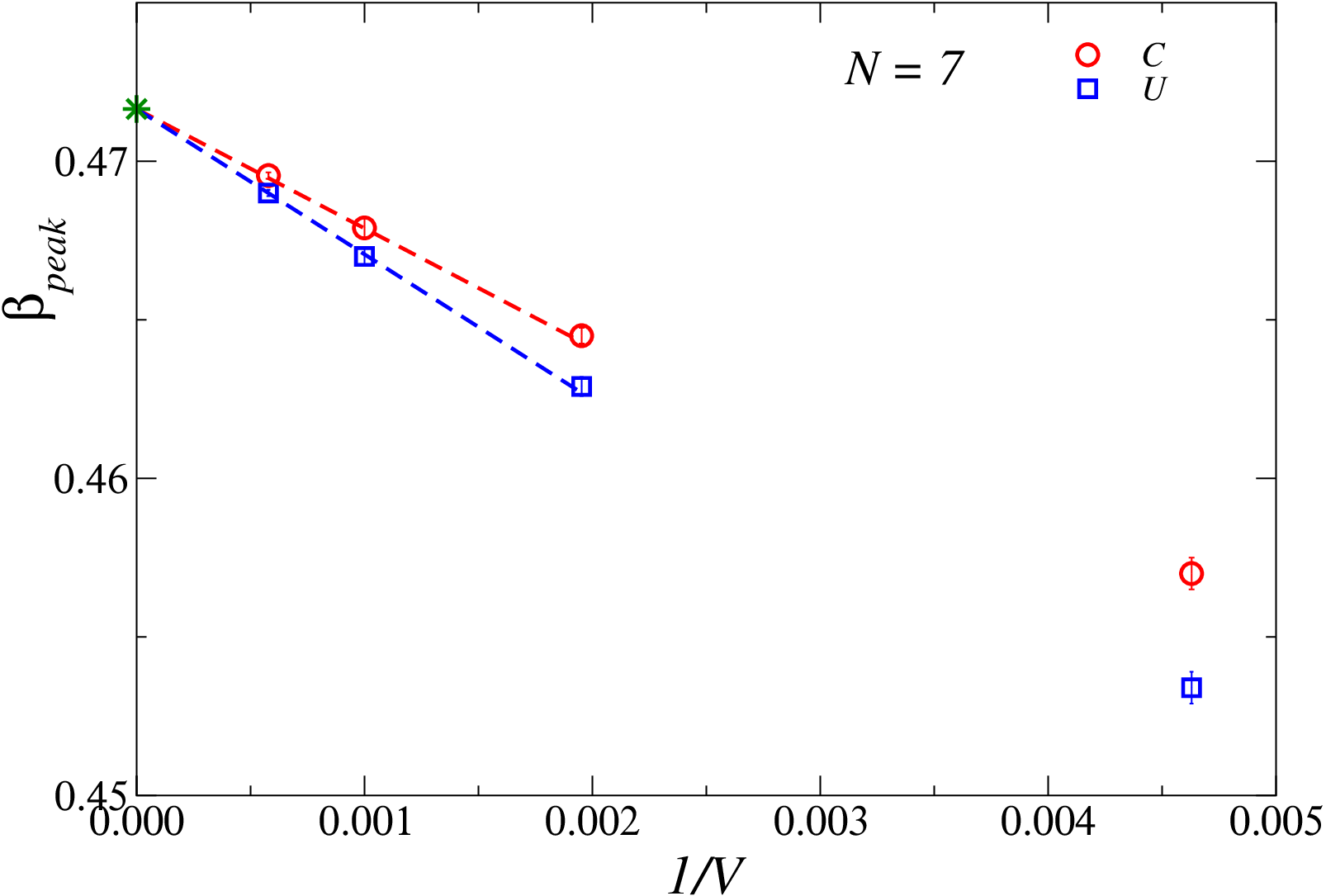}
\end{center}
\includegraphics[width=0.5\textwidth, keepaspectratio]{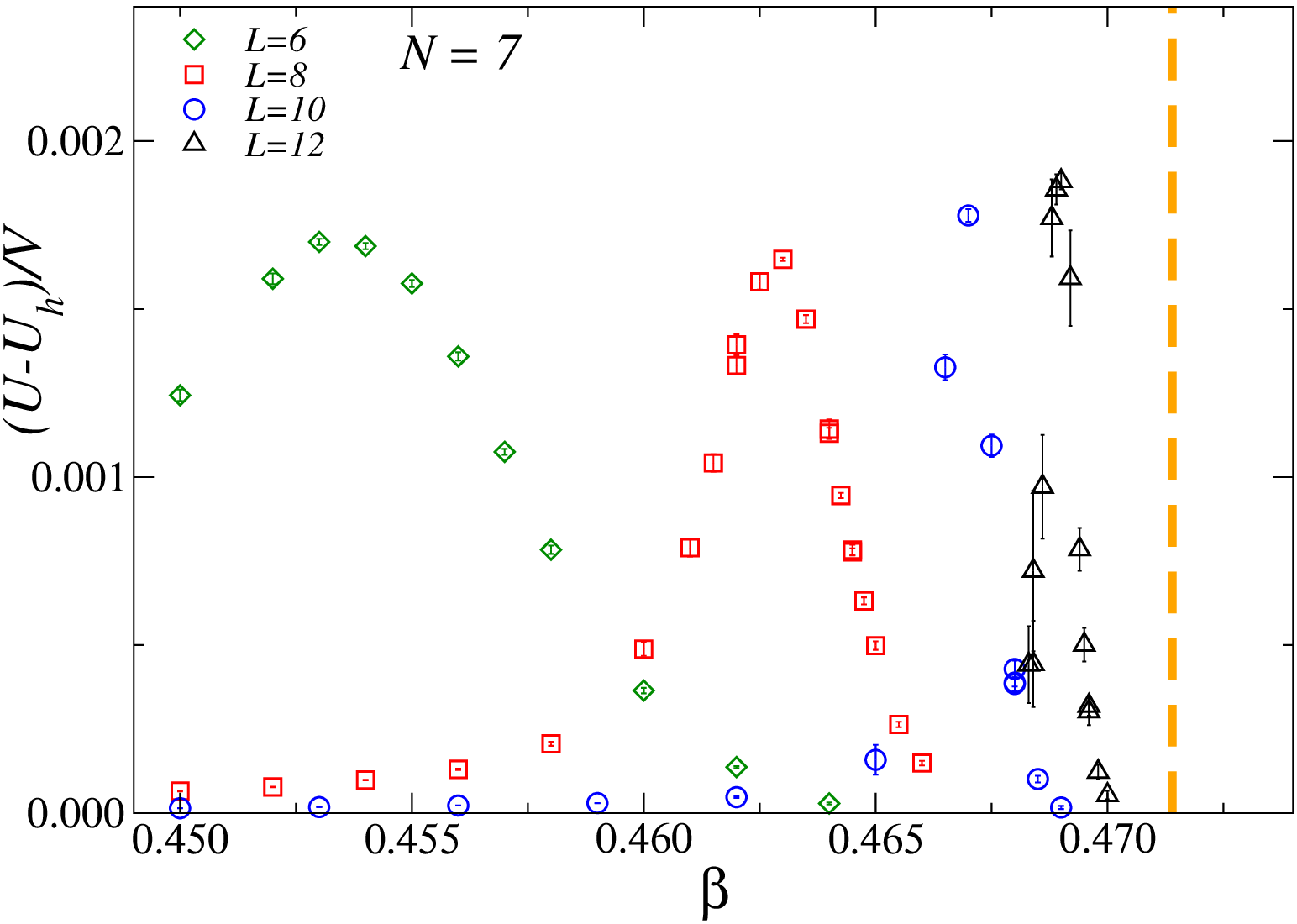}
\includegraphics[width=0.5\textwidth, keepaspectratio]{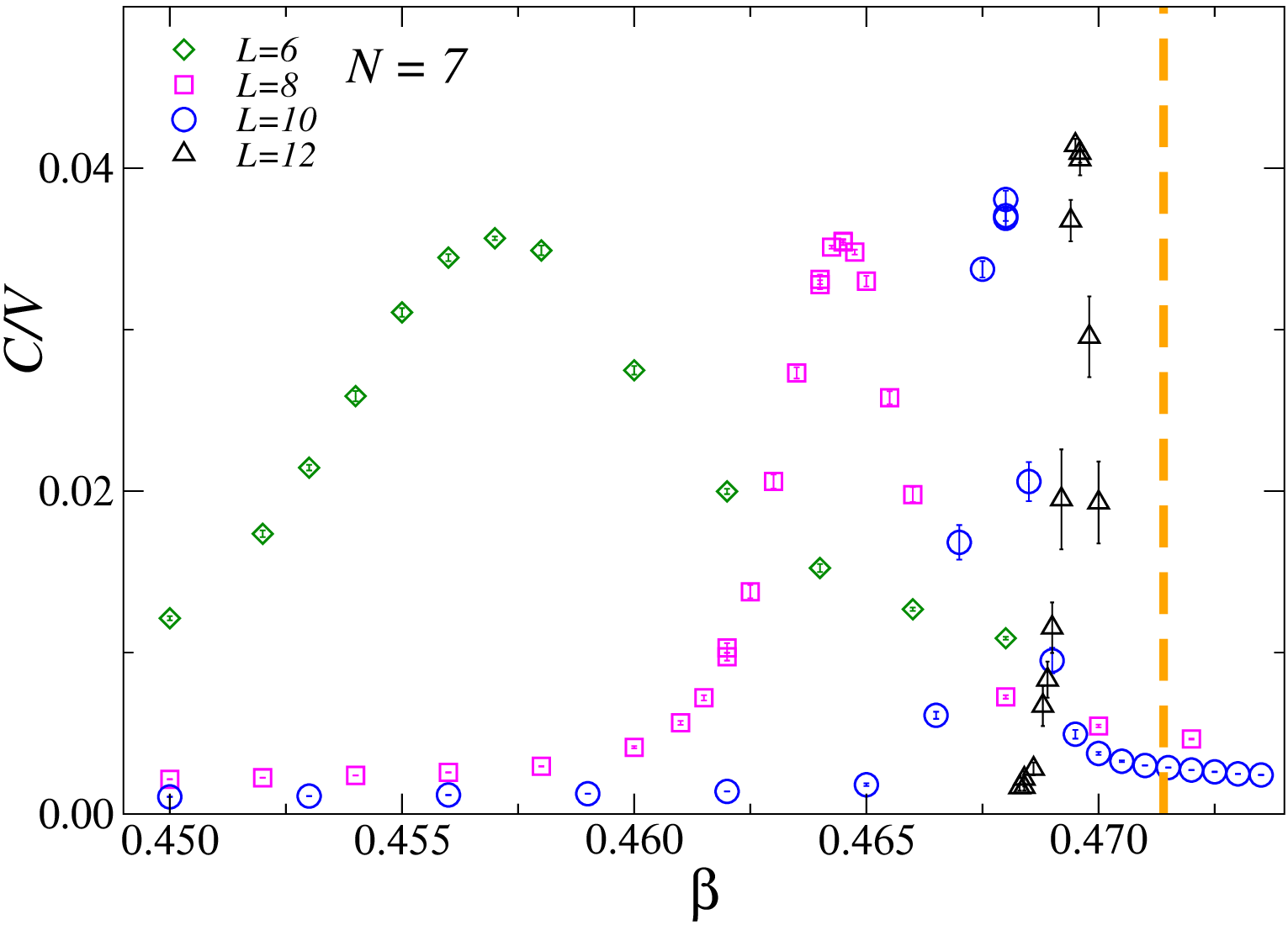}
\caption{Estimates of the Binder cumulant $U$ (bottom left), of the
  specific heat (bottom right), and of the positions $\beta_{{\rm
      max},C}$ and $\beta_{{\rm max},U}$ of the maxima (top) for
  $N=7$. In the top panel we also report the extrapolations that
  provide an estimate of $\beta_c$.  }
\label{n7}
\end{figure}

\begin{figure}[tbp]
\begin{center}
\includegraphics[width=0.45\textwidth,keepaspectratio,angle=0]{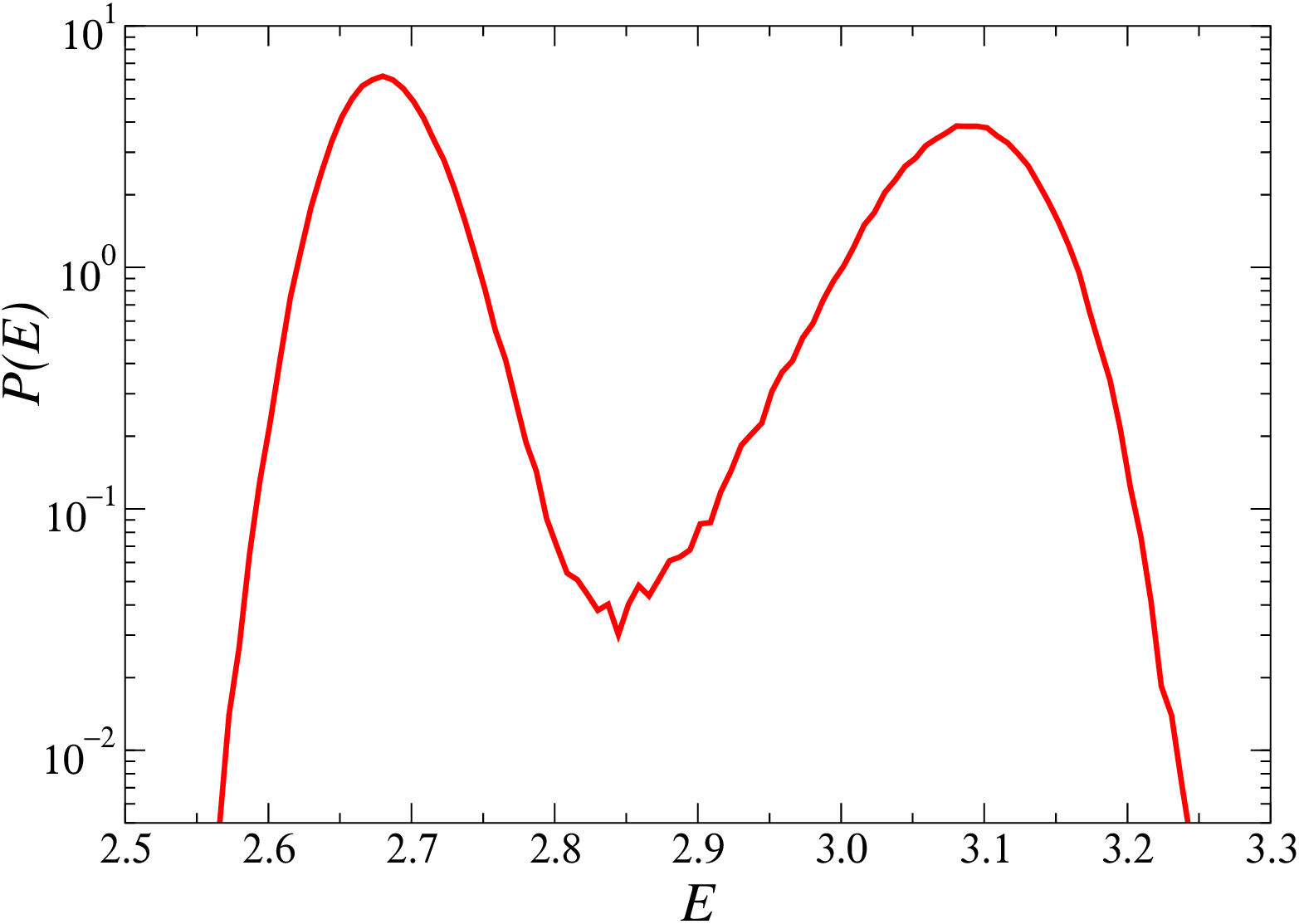}
\includegraphics[width=0.45\textwidth,keepaspectratio,angle=0]{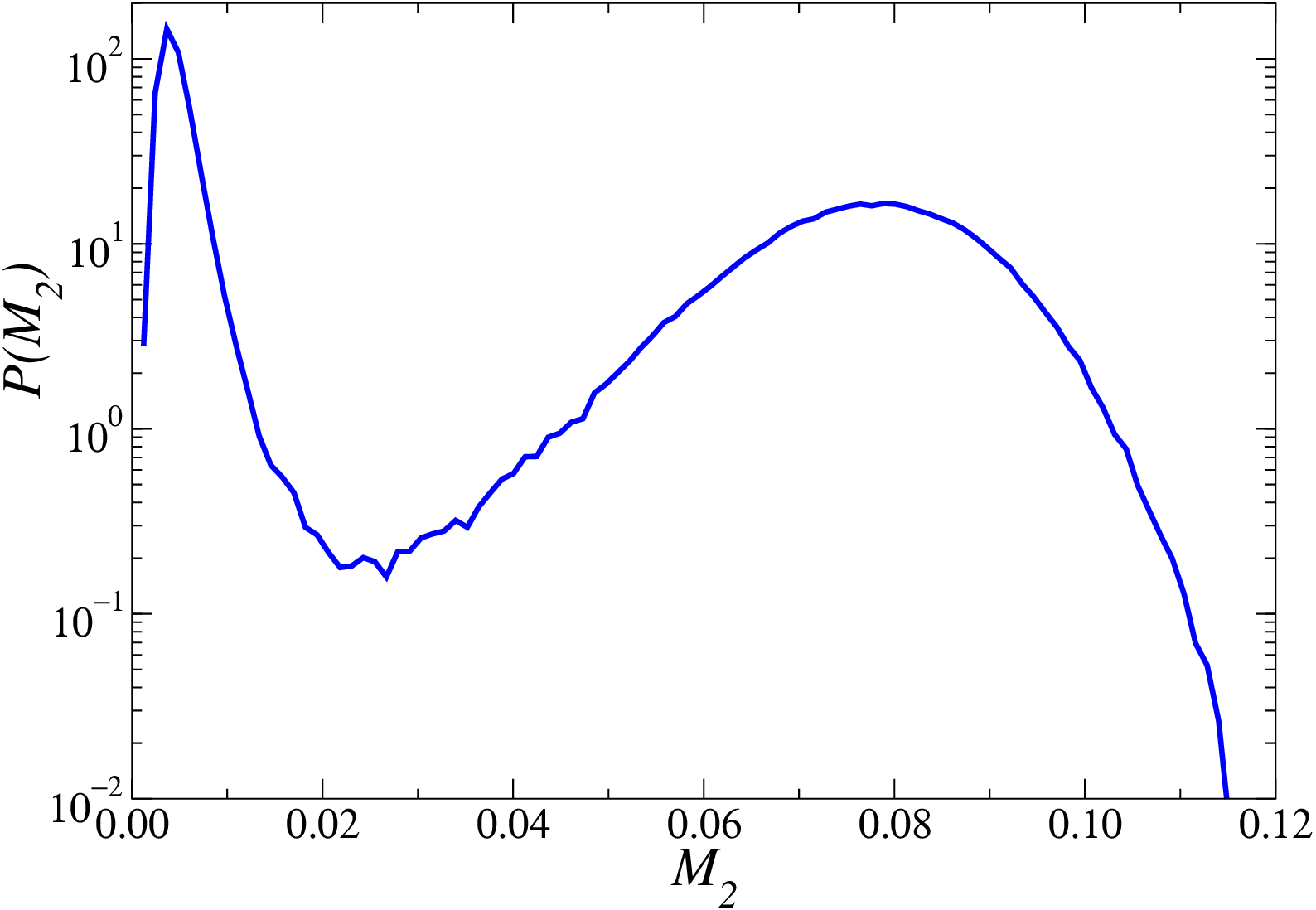}
\end{center}
\caption{Distributions of the energy $E$ (left) and of $M_2 =
  \mu_2/V^2$ (right), where $\mu_2$ is defined in
  Eq.~(\ref{binderdef}) and $V$ is the volume.  Results for $N=7$,
  $L=12$, $\beta = 0.4693\approx \beta_{C,\rm max}(L)$.  }
\label{istogrammi-N7}
\end{figure}

In Fig.~\ref{n7} we report results for $N=7$.  We plot the ratios
$C/V$ and $(U-U_h)/V$, where $U_h=(N^2+1)/(N^2-1)$ is the
high-temperature (HT) value of the Binder parameter. As we discussed
in \cite{PV-19}, the subtracted term, although asymptotically
irrelevant, allows us to take somehow into account the corrections of
order $V^{-1}$ to the asymptotic behavior of $U$, cf. 
Eq.~(\ref{CmaxUmax}).  The reported
results are consistent with $U,C\sim V$, and therefore provide clear
evidence for a first-order transition.  The extrapolations of
$\beta_{{\rm max},C}(L)$ and of $\beta_{{\rm max},U}(L)$ allow us to
estimate $\beta_c$. The two extrapolations give consistent results: we
estimate $\beta_c=0.4714(5)$.  The first-order nature of the
transition is also confirmed by the two-peak structure of the
distributions of $E$ and of the square of the local order parameter
$\mu_2/V^2$ [$\mu_2$ is defined in Eq.~(\ref{binderdef})]; see
Fig.~\ref{istogrammi-N7} for results for $L=12$ and $\beta \approx
\beta_{{\rm max},C}(L)$.  If $P_{\rm max}$ is the maximum value of the
distribution of $E$ and $P_{\rm min}$ is the minimum value in the
valley between the two maxima, we observe that $P_{\rm min}/P_{\rm
  max} \approx 10^{-2}$ for $L=12$, which indicates a relatively
strong transition. Since this ratio is supposed to scale as $e^{-\beta
  \sigma L^2}$, where $\sigma$ is the surface tension, assuming a
prefactor of order one, we predict the ratio $P_{\rm min}/P_{\rm max}$
to be of order $10^{-4}$ for $L=16$, which indicates that a standard
local algorithm is not able to sample correctly both phases for $L\ge
16$ (our runs consist in O$(10^k)$ lattice sweeps with $k\approx
6$-7).  For this reason we have only results with $L\le 12$.

\begin{figure}[tbp]
\begin{center}
\includegraphics*[width=0.5\textwidth, keepaspectratio]{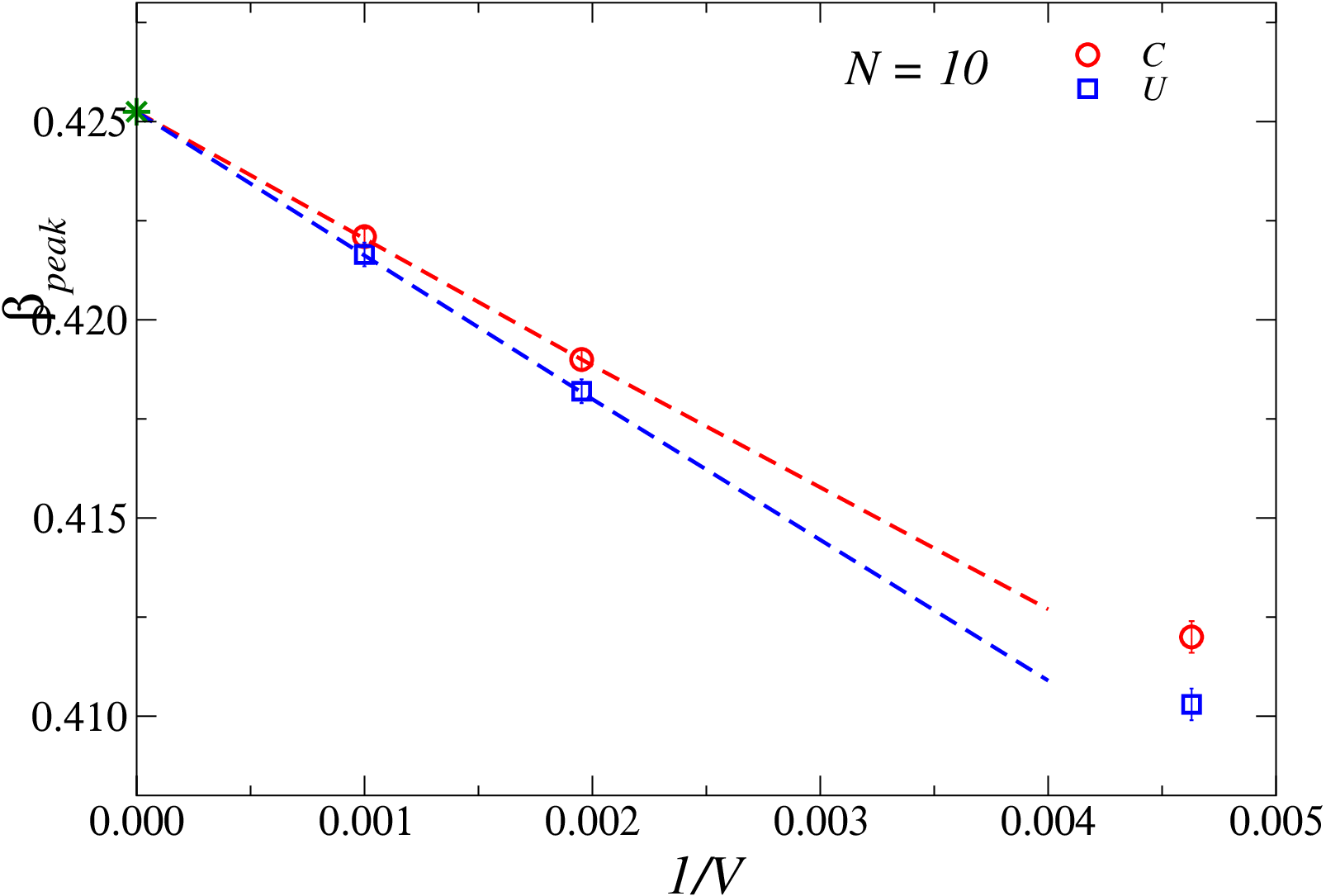}
\end{center}
\includegraphics*[width=0.5\textwidth, keepaspectratio]{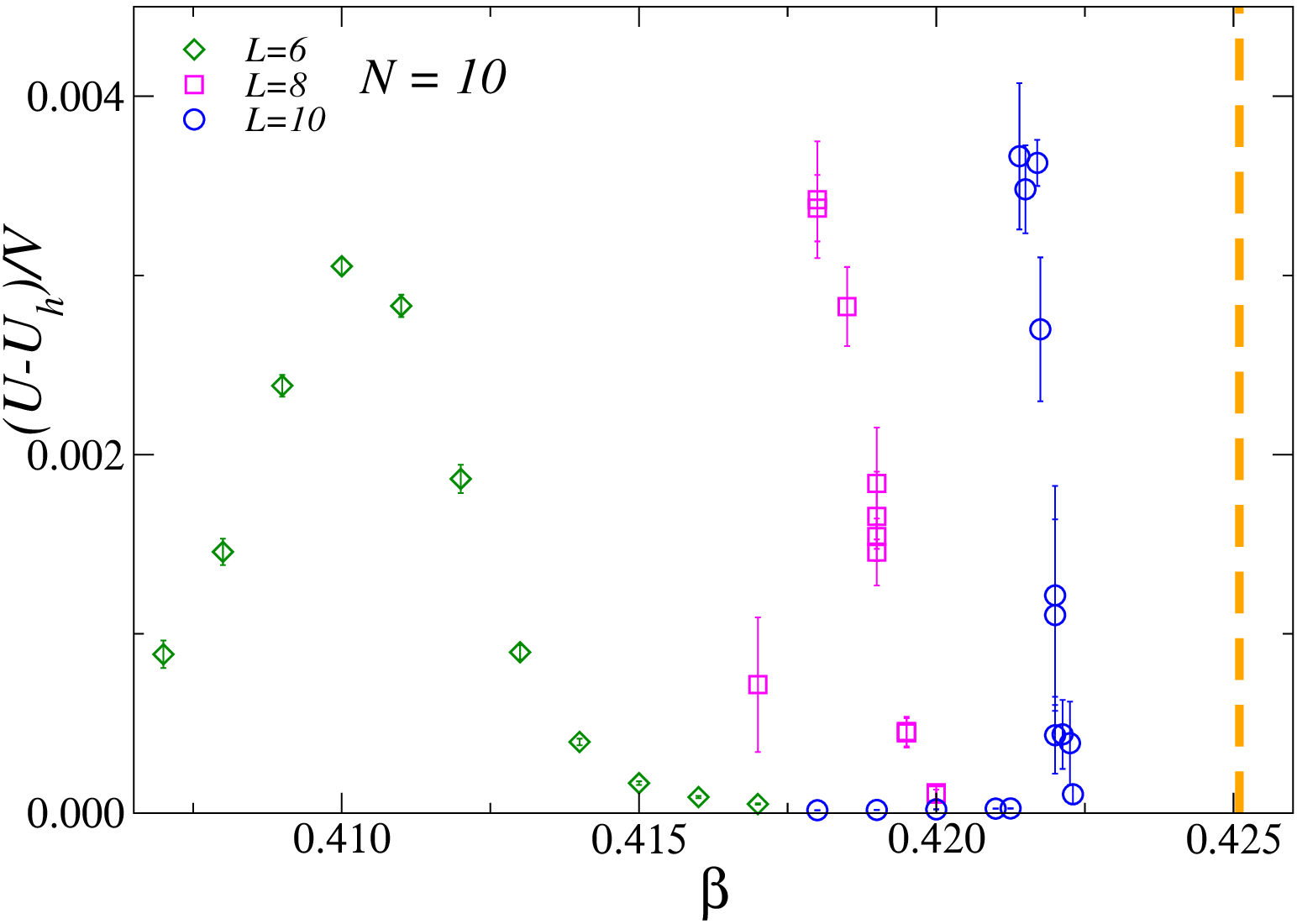}
\includegraphics*[width=0.5\textwidth, keepaspectratio]{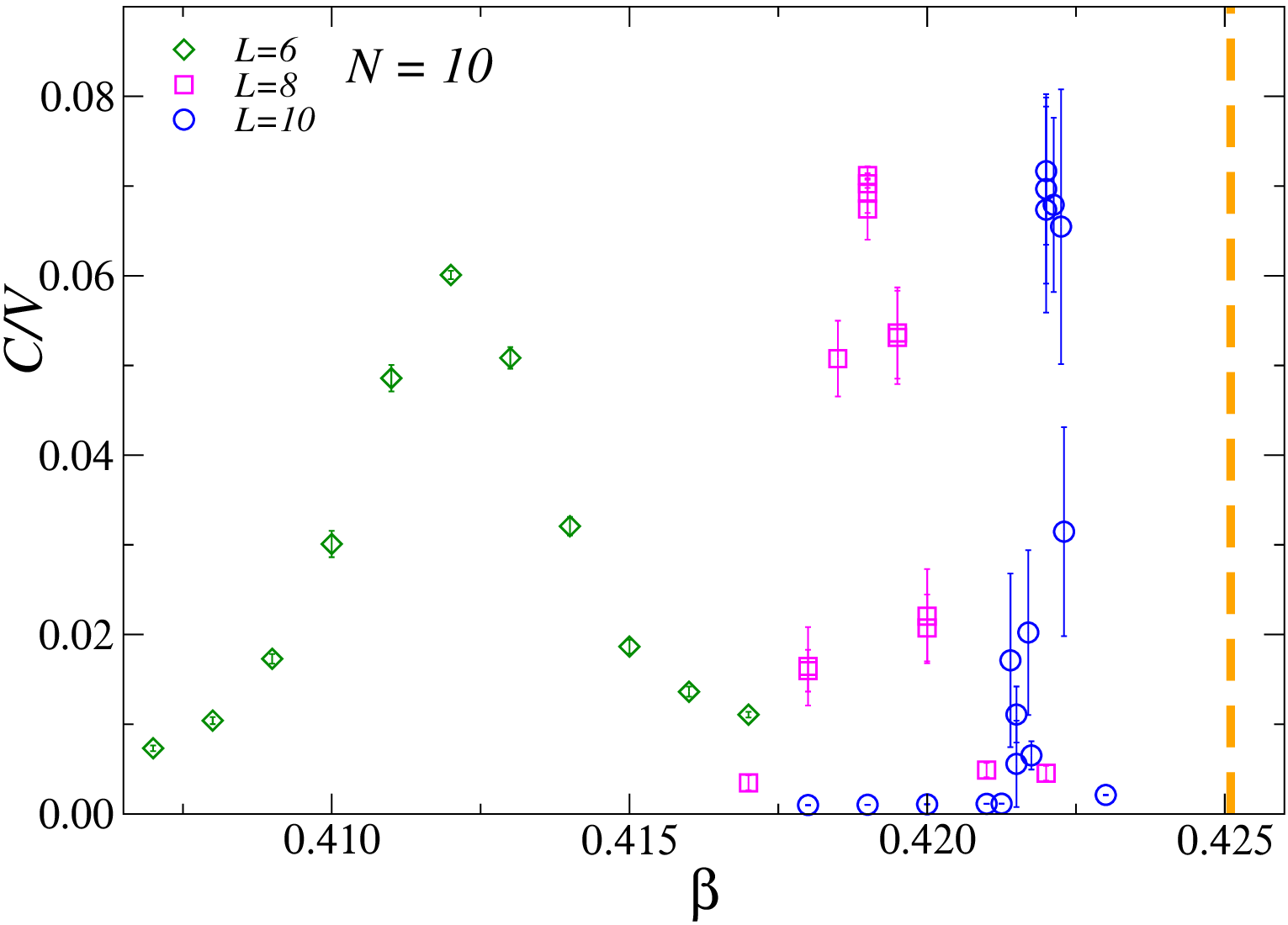}
\caption{Estimates of the Binder cumulant $U$ (bottom left), of the
  specific heat (bottom right), and of the positions $\beta_{{\rm
      max},C}$ and $\beta_{{\rm max},U}$ of the maxima (top) for
  $N=10$. In the top panel we also report the extrapolations that
  provide an estimate of $\beta_c$.  }
\label{n10}
\end{figure}

\begin{figure}[tbp]
\begin{center}
\includegraphics*[width=0.5\textwidth, keepaspectratio]{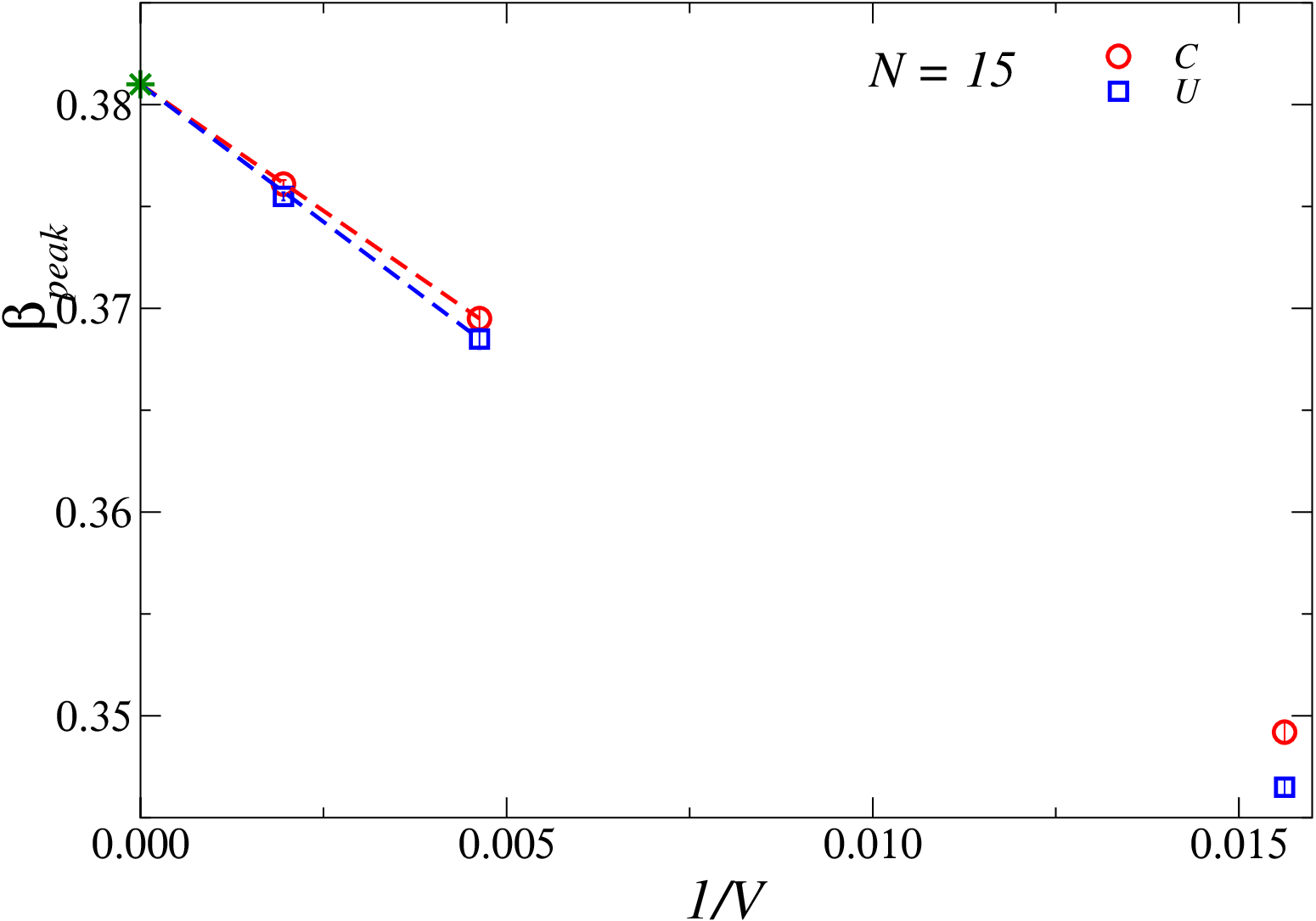}
\end{center}
\includegraphics*[width=0.5\textwidth, keepaspectratio]{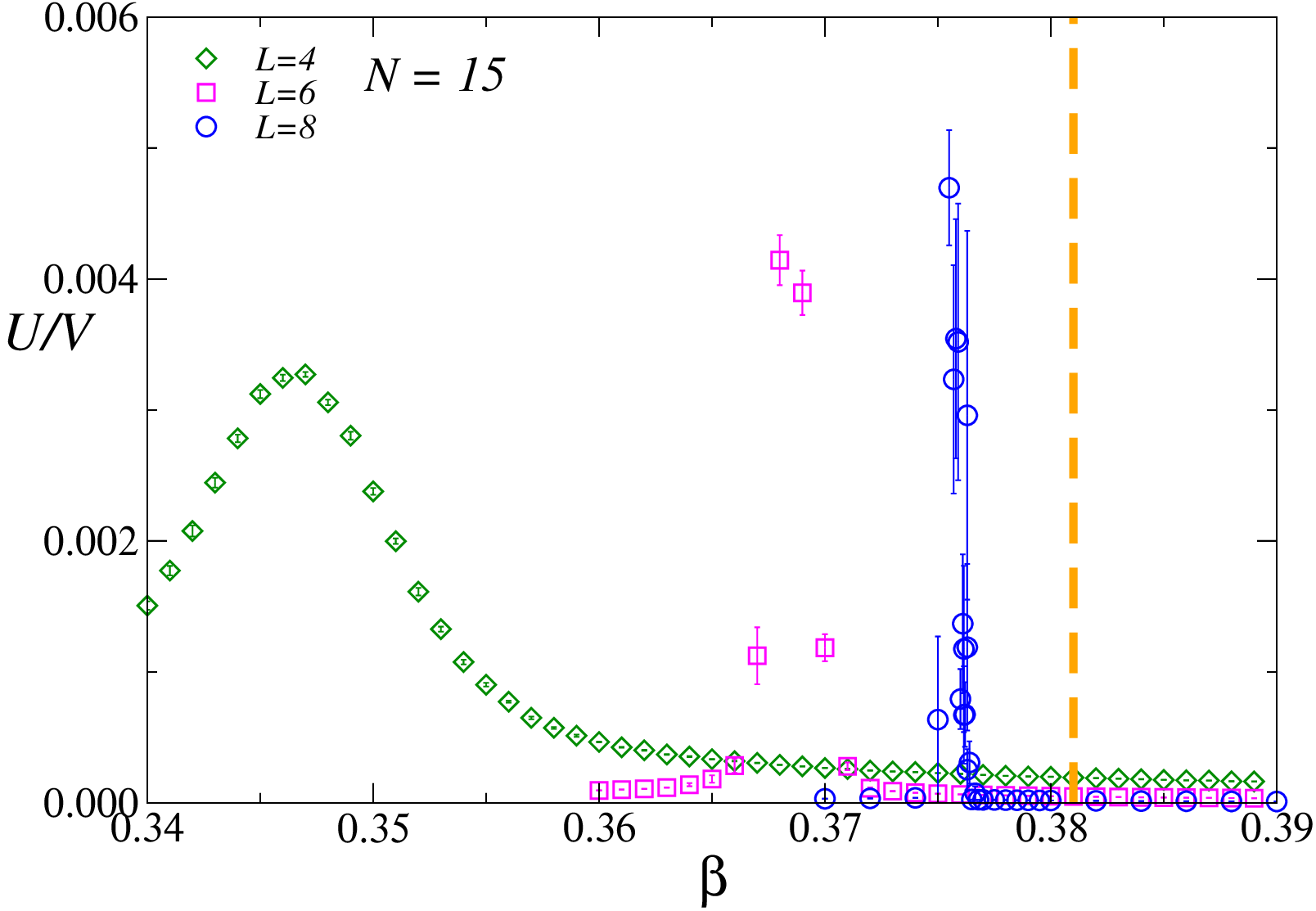}
\includegraphics*[width=0.5\textwidth, keepaspectratio]{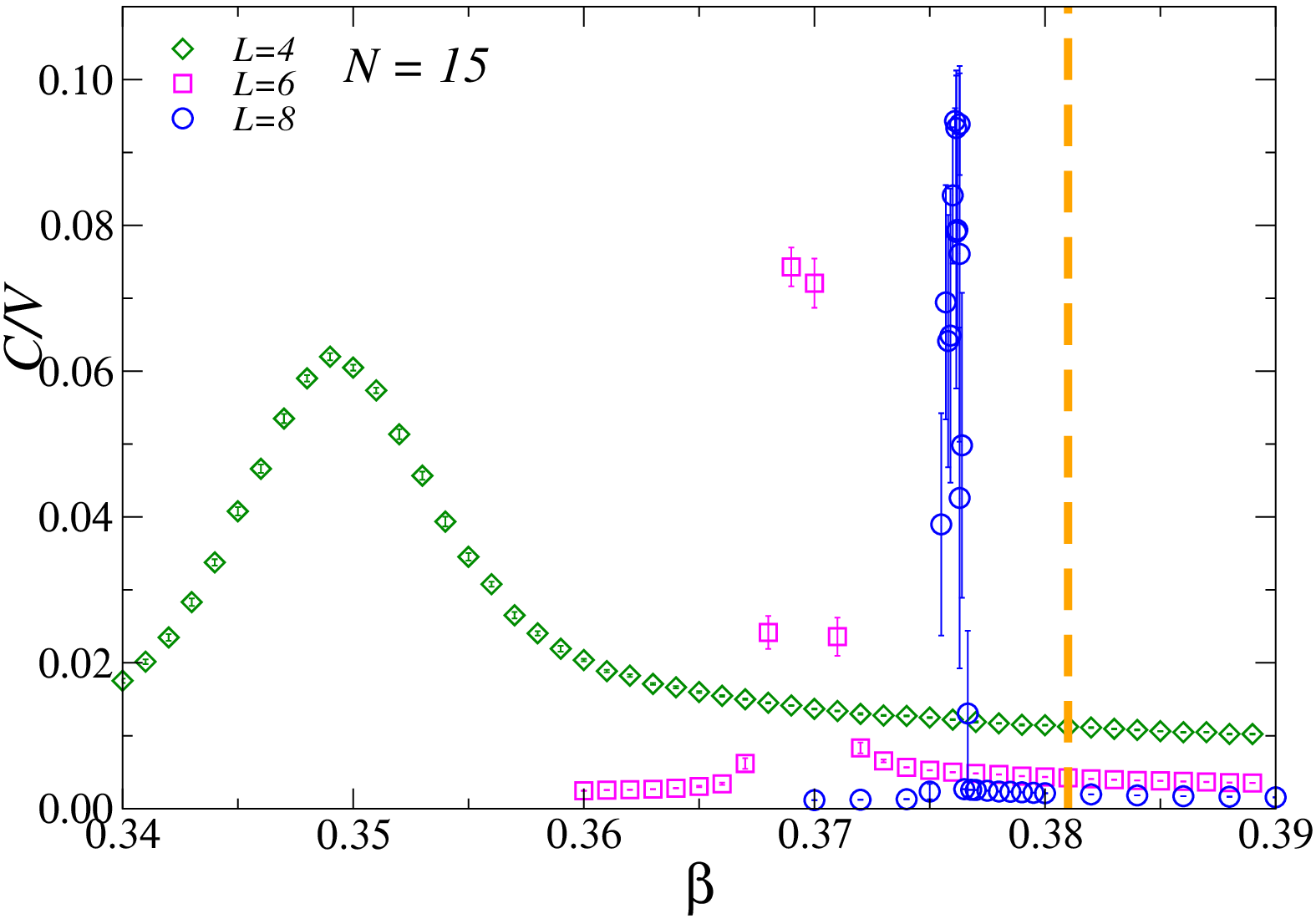}
\caption{Estimates of the Binder cumulant $U$ (bottom left), of the
  specific heat (bottom right), and of the positions $\beta_{{\rm
      max},C}$ and $\beta_{{\rm max},U}$ of the maxima (top) for
  $N=15$. In the top panel we also report the extrapolations that
  provide an estimate of $\beta_c$.  }
\label{n15}
\end{figure}

\begin{figure}[tbp]
\begin{center}
\includegraphics*[width=0.5\textwidth, keepaspectratio]{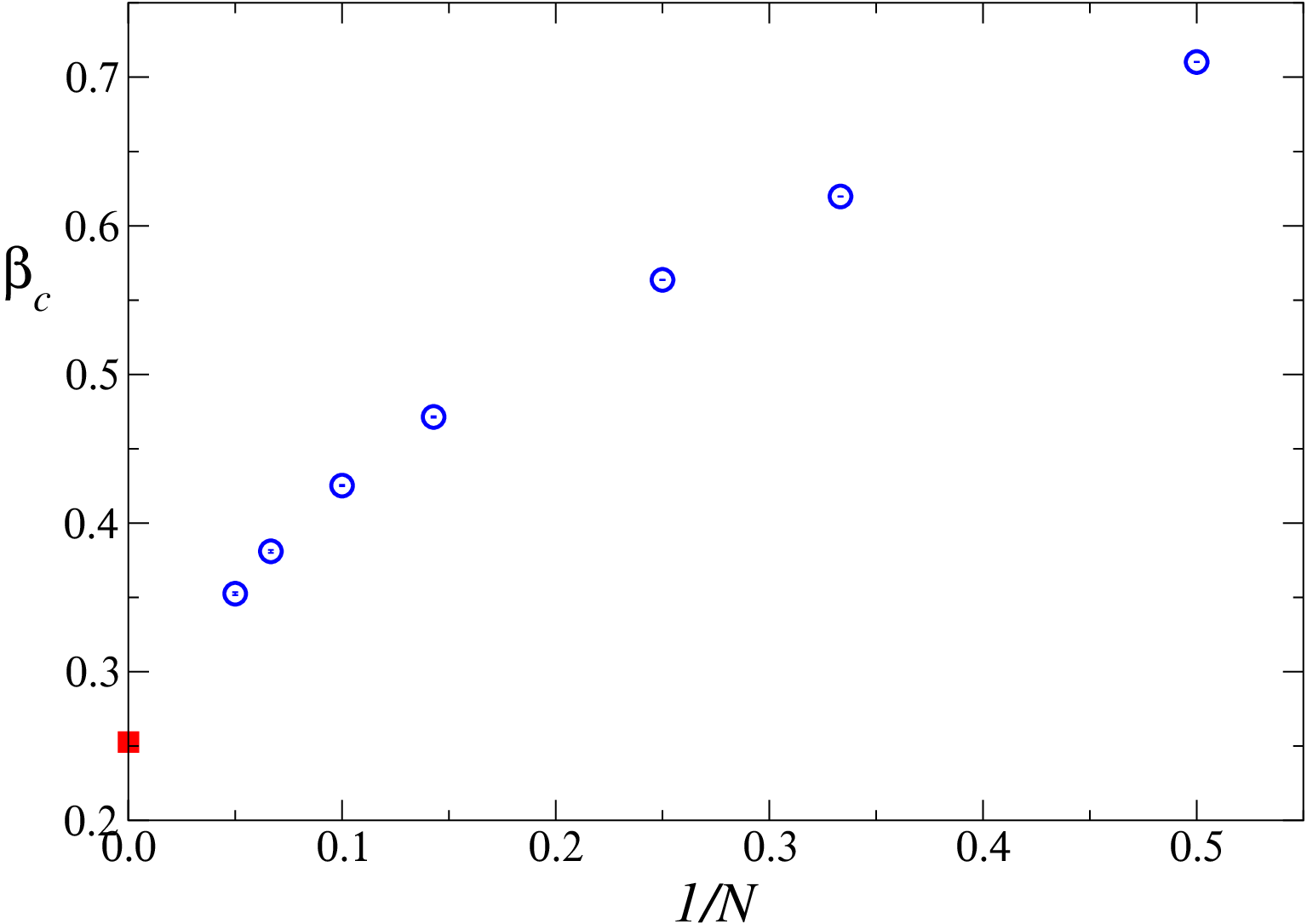}
\end{center}
\caption{ Plot of the available estimates of $\beta_c$ versus $1/N$.
  For $1/N=0$, we report $\beta_{c,\infty}\approx 0.25273$, which is
  probably a lower bound on the correct critical temperature for
  $N=\infty$ (see Sec.~\ref{largeN-sol} for a discussion).  }
\label{betac}
\end{figure}

Similar results hold for $N=10$ and 15, see Figs.~\ref{n10} and
\ref{n15}.  We observe a first-order transition at $\beta_c=0.4253(5)$
for $N=10$ and at $\beta_c=0.381(1)$ for $N=15$.  The transition
becomes stronger as $N$ increases: the ratio $P_{\rm min}/P_{\rm max}$
at fixed $L$ decreases significantly as a function of $N$---therefore,
the surface tension that parametrizes the interface free energy
increases---limiting us to smaller and smaller values of $L$.  For
$N=20$ we are not  able to go beyond $L=6$ and therefore, we cannot
make a quantitative study of the transition.  We only roughly estimate
the transition temperature, $\beta_c\approx 0.353$.  In
Fig.~\ref{betac} we plot the estimates of $\beta_c$ versus $1/N$ (we
also include the results of \cite{PV-19}), together with the large-$N$
estimate of \cite{PV-19}, $\beta_{c,\infty}=0.25273...$, which is
probably a lower bound to the correct value (this is discussed in
Sec.~\ref{largeN-sol}).

\begin{table}
\caption{Estimates at the transition point of the latent heat
  $\Delta_h$ and of the correlation lengths obtained from correlations
  of $Q$ [$\xi$, Eq.~(\ref{xidefpb})], from the Polyakov loop
  [$\xi_P$, Eq.~(\ref{Polyakov-exp})] and from the gauge correlations
  $G_V(x)$ [$\xi_z$, Eq.~(\ref{GV-exp})]. The suffix HT (LT) refers to
  the high-temperature (low-temperature, resp.) phase.  The results
  for $\Delta_h$, $\xi_{HT}$ and $\xi_{z,HT}$ have been obtained on a
  lattice of size $L=24$, those for $\xi_{z,LT}$ on a lattice of size
  $L=48$ ($N=7$) or $L=40$ ($N=10,20$). }
\label{table-betac}
\begin{center}
\begin{tabular}{cccccc}
\hline\hline
$N$ & $\Delta_h$ & $\xi_{HT}$ & $\xi_{P,LT}$ & $\xi_{z,HT}$  &  $\xi_{z,LT}$ \\
\hline
7   &  0.4066(3) & 1.710(6)   & 4.17(2) &  1.46(1) &  4.01(2)   \\
10  &  0.5397(2) & 1.169(5)   &         &  1.28(1) &  5.22(5)   \\
20  &  0.5806(2) & 0.700(3)   & 6.71(2) &  1.03(1) &  6.95(20)  \\
\hline\hline
\end{tabular}
\end{center}
\end{table}

\begin{figure}[tbp]
\begin{center}
\includegraphics[width=0.5\textwidth, keepaspectratio,angle=0]{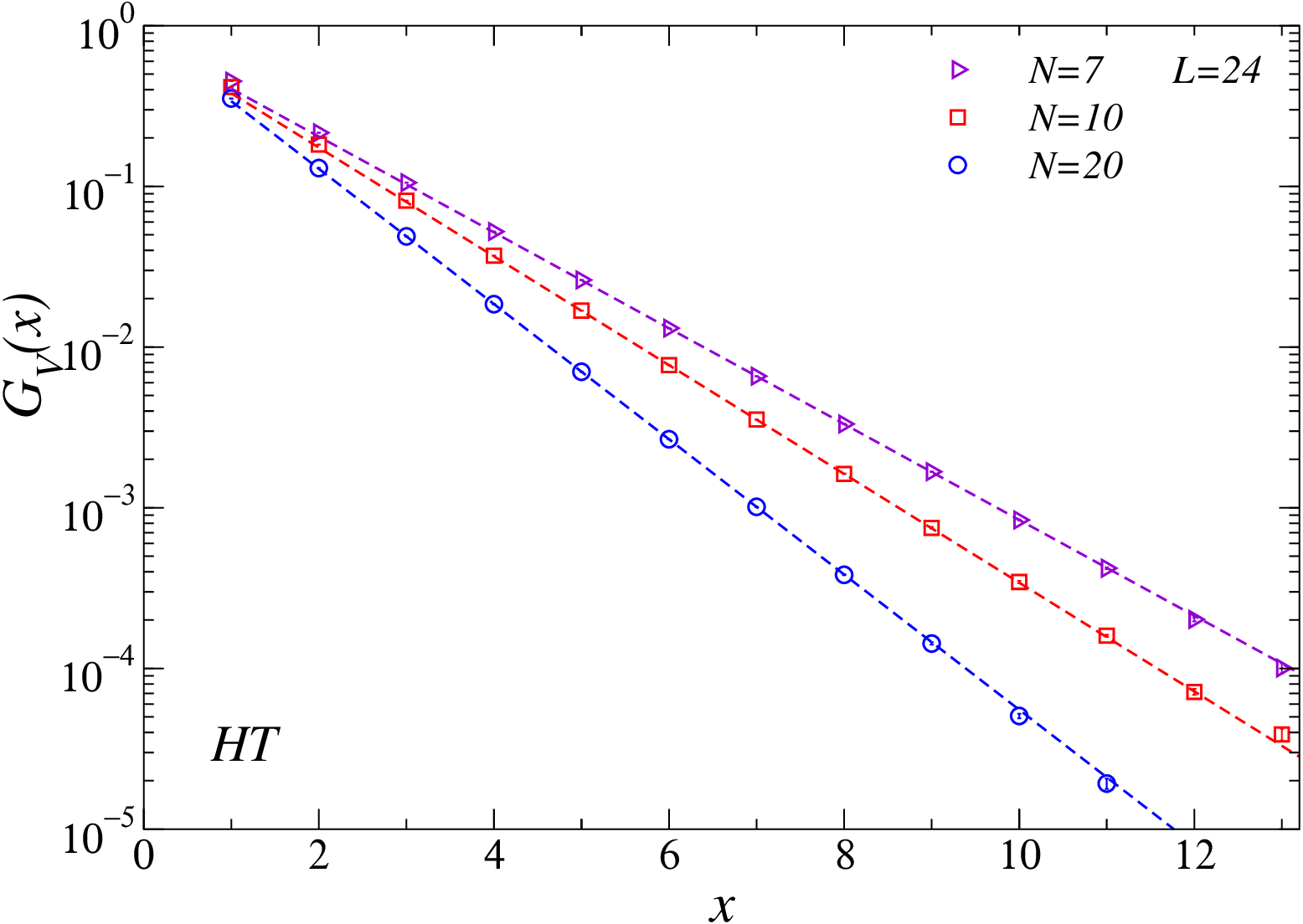}
\includegraphics[width=0.5\textwidth, keepaspectratio,angle=0]{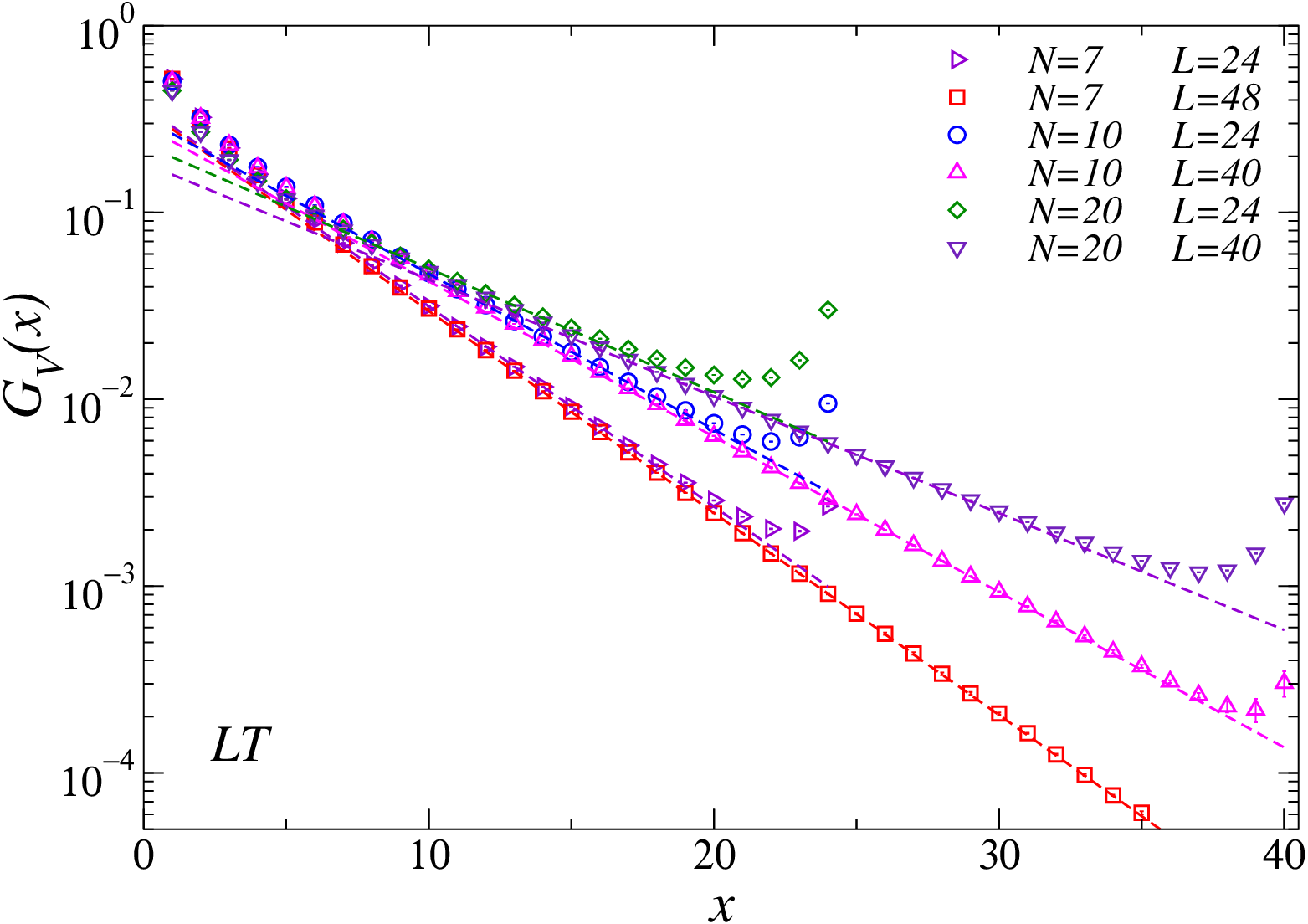}
\end{center}
\caption{ Plot of the correlation function $G_V(x)$ versus $x$, in the
  low-temperature (LT, bottom) and in the high-temperature phase (HT,
  top).  Results at $\beta\approx\beta_c$ obtained starting the
  simulation from an ordered (LT) or disordered (HT) configuration.
  The dashed lines going through the points are fits to
  Eq.~(\ref{GV-exp}).  }
\label{correl-betac}
\end{figure}

To analyze the behavior of the system in the two phases at $\beta
\approx \beta_c$, we proceed as follows.  We fix $\beta$ to our
estimate of $\beta_c$ and perform two runs, which start from a
disordered and an ordered configuration, respectively. If $L$ is large
enough (as we discussed, it is enough to take $L\gtrsim 16$), during
the simulation there are no phase swaps and therefore, we are able to
determine the average values of the different observables in the two
phases.  Using this method, we have estimated the average energy in
the two phases and the corresponding latent heat. The results for
$L=24$, reported in Table~\ref{table-betac}, are consistent with those
that can be obtained from the behavior of the specific heat maximum,
see Eq.~(\ref{CmaxUmax}).  We have not attempted an infinite-volume
extrapolatiom, but comparison with results for smaller values of $L$
indicates that size deviations are significantly less than 1\%. The
data show that $\Delta_h$ increases as $N$ increases: the first-order
transition becomes stronger in the large-$N$ limit.

Similar conclusions are reached from the analysis of the correlations
of the order parameter.  In the low-temperature (LT) phase, $\xi$
computed from the $Q$ correlations, see Eq.~(\ref{xidefpb}), increases
with $L$ for any $N$. This is of course expected, as the order
parameter $Q$ condenses in the LT phase. On the other hand, in the HT
phase, $\xi$ decreases with increasing $N$ and is always of order
one. Apparently, correlations do not develop on the HT side of the
transition as $N$ increases. This is obviously in contrast with the
idea that the transition becomes continuous for large values of
$N$.  In this case, one would expect $\xi$ to increase with $N$,
becoming of order $L$ for $N$ large enough, as expected in the
vicinity of a critical transition.

To analyze the behavior of the gauge-field dependent observables, we
first consider the Polyakov loop. Such a quantity is generically
expected to decay exponentially with the system size \cite{PV-19},
i.e.,
\begin{equation}
  P(L) = A_P e^{-L/\xi_P},
\label{Polyakov-exp}
\end{equation}
with an appropriate correlation length $\xi_P$.  In the HT phase, even
for $L$ as small as 12, the Polyakov loop is negligible within errors
(they are of the order of $10^{-5}$). If $A_P$ is a constant of order
1, this implies that $\xi_P$ is approximately 1 or smaller ($e^{-12} =
6\cdot 10^{-6}$). In the HT phase, therefore, also gauge modes are
essentially uncorrelated. In the LT phase, we can estimate $\xi_P$
considering data in the range $16\le L \le 40$. The results show that
$\xi_P$ increases with $N$: gauge correlations become larger in the
large-$N$ limit. Finally, we have considered the correlation function
$G_V(x)$.  As shown in Fig.~\ref{correl-betac}, the curves behave
quite precisely as exponentials, i.e., they are well fitted by
\begin{equation}
G_V(x) = A e^{-x/\xi_z}.
\label{GV-exp}
\end{equation}
If we fit the numerical results to Eq.~(\ref{GV-exp}), we obtain
estimates of $\xi_z$ (they are reported in Table~\ref{table-betac}),
that are consistent with a very simple scenario: gauge correlations
are always negligible in the HT phase---for any $N$ and, therefore,
also for $N\to \infty$---while in the LT phase they increase with $N$,
leaving open the possibility that $\xi_z$ becomes infinite for
$N\to\infty$.

\begin{table}
\caption{Estimates of the different correlation lengths as a function
  of $N$.  In the second and third columns we report results obtained
  at $\beta = 0.25273 \approx \beta_{c,\infty}$, in the fourth and
  fifth column results for $\beta = 0.8$. The HT results are obtained
  on lattices of size $L=16$; the LT estimates of $\xi_z$ on lattices
  with $L=24$ ($N=10,20$) and $L=32$ ($N=50,100$). The Polyakov
  correlation length is a fit of results with $16\le L \le 32$.}
\label{table:phases}
\begin{center}
\begin{tabular}{ccccc}
\hline\hline
$N$  &   \multicolumn{2}{c}{$\beta=0.25273$} & 
         \multicolumn{2}{c}{$\beta=0.8$}  \\
     &   $\xi$  & $\xi_z$  &   $\xi_P$  & $\xi_z$ \\
\hline
10   &   0.275(9) & 0.70(1)  &   & 22.8(3) \\
20   &   0.285(5) & 0.71(1)  &   & 47(3) \\
50   &   0.316(2) & 0.72(1)  & 116.2(1)  & 116(10) \\
100  &   0.437(1) & 0.87(2)  & 233.4(1)  & 225(20) \\
\hline\hline
\end{tabular}
\end{center}
\end{table}

\begin{figure}[tbp]
\begin{center}
\includegraphics*[width=0.5\textwidth,keepaspectratio,angle=0]{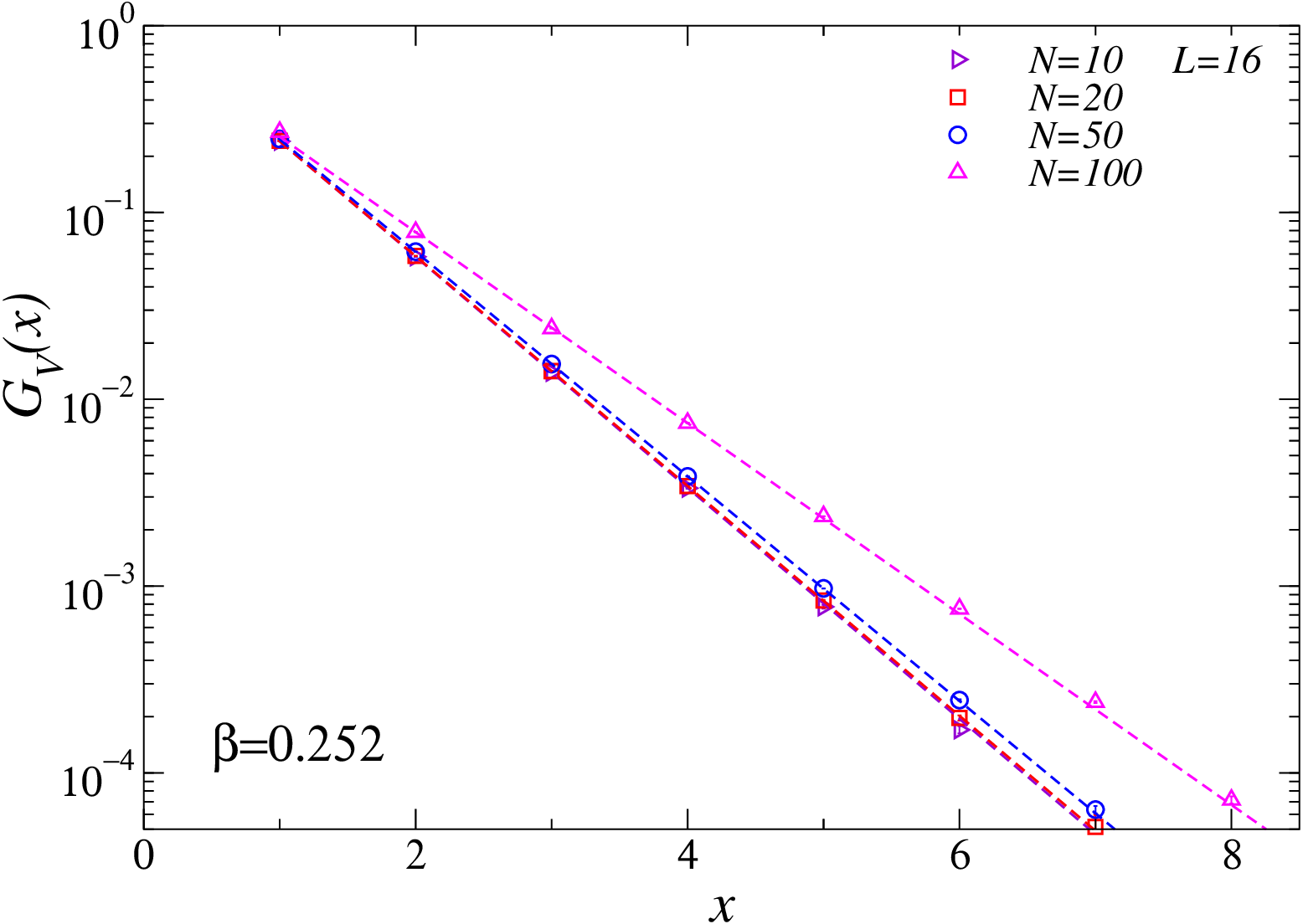}
\includegraphics*[width=0.5\textwidth,keepaspectratio,angle=0]{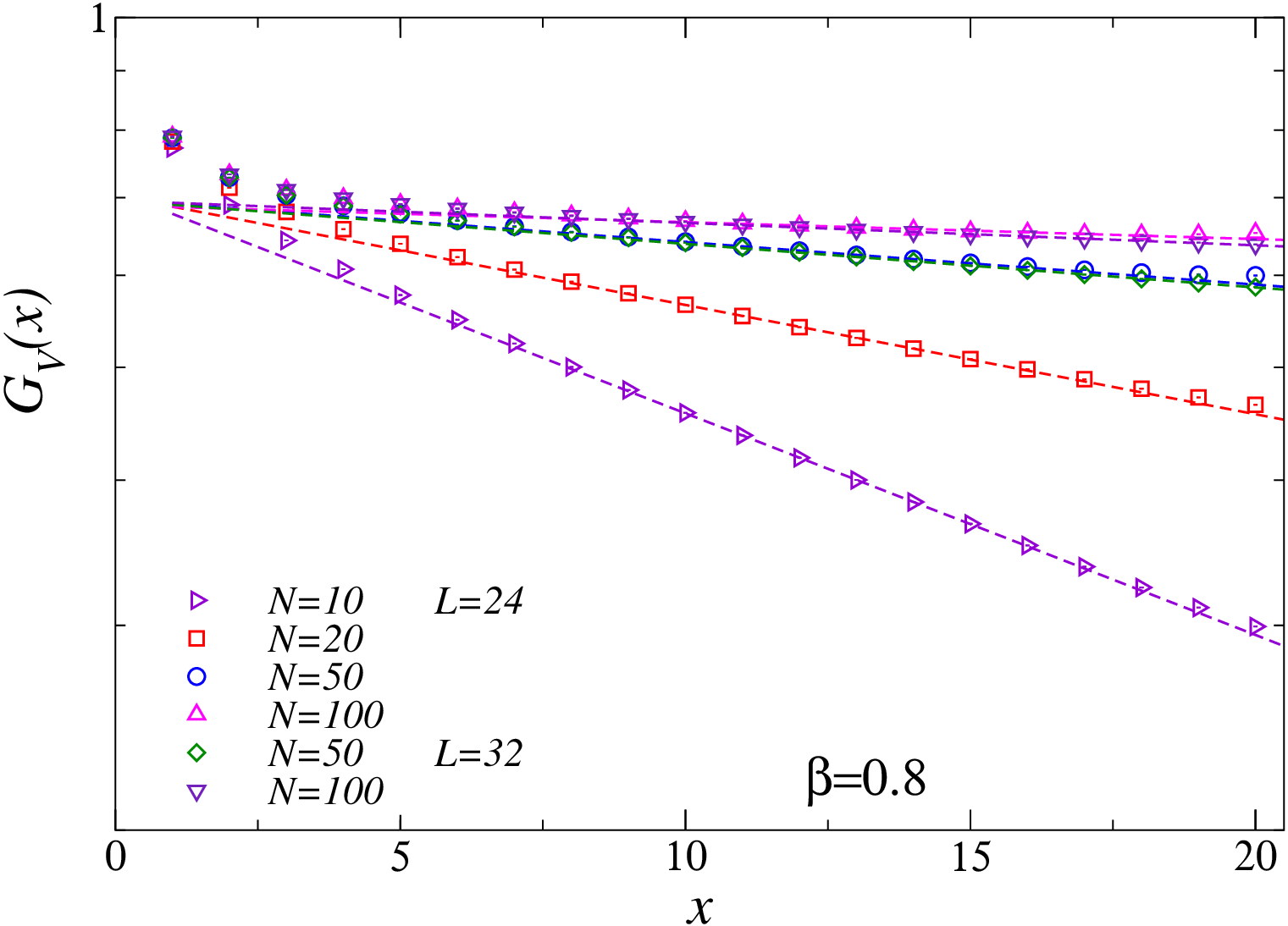}
\end{center}
\caption{ Plot of the correlation function $G_V(x)$ versus $x$, in the
  LT phase ($\beta = 0.8$, bottom) and in the HT phase ($\beta =
  0.25273$, top), for different values of $N$.  The dashed lines going
  through the points are fits to Eq.~(\ref{GV-exp}).  }
\label{correl-fasi}
\end{figure}

Summarizing, we have shown that, at least up to $N=20$, the model
undergoes a first-order transition that apparently becomes stronger as
$N$ increases. The transition separates two phases. The HT phase is
disordered: both correlations of the order parameter $Q_{\bm x}$ and
gauge correlations decay very rapidly, with a typical length scale of
a lattice spacing.  Moreover, the correlation lengths $\xi$ and
$\xi_x$ apparently decrease as $N$ increases. In the LT phase the
order parameter $Q_{\bm x}$ condenses and $\xi\sim L$.  Gauge
correlations are always massive, with a corresponding correlation
length that increases with $N$. These results allow us to formulate a
simple scenario for the behavior in the large-$N$ limit.  We expect
the $N=\infty$ transition to be of first order. For $\beta<\beta_c$
gauge and $Q$ correlations are always massive, while for $\beta >
\beta_c$ both gauge-dependent vector correlations and gauge-invariant
$Q$ correlations are massless: in the infinite-volume limit both $\xi$
and $\xi_z$ are infinite for $N=\infty$.

\subsection{Phase behavior}  \label{sec2.3}

Here we wish to provide additional support to the scenario discussed
in the previous section, determining how $\xi$, $\xi_P$, and $\xi_z$
vary as a function of $N$ for two different fixed values of
$\beta$. First, we consider $\beta = 0.25273 \approx
\beta_{c,\infty}$, where $\beta_{c,\infty}$ is the transition point
predicted by the large-$N$ analysis of \cite{PV-19}.  If the standard
large-$N$ analysis is correct, these runs should allow us to determine
the behavior of the model on the HT side of the large-$N$ transition
point.  We will also perform runs at $\beta = 0.8$, deep in the LT
phase.  We will consider four different values of $N$, $N=10,20,50,$
and 100.

Let us first consider the runs in the HT phase, at $\beta = 0.25273$.
Results for $L=16$ are reported in Table~\ref{table:phases}. For all
values of $N$, $\xi$ is very small, consistent with a correlation of
the order of at most one lattice spacing in the limit $N\to \infty$. A
diverging correlation length for $N=\infty$ is clearly not consistent
with the data. In Fig.~\ref{correl-fasi} (top panel) we report the
correlation function $G_V(x)$.  It decays very rapidly with $x$ with a
correlation length $\xi_z$ that is little dependent on $N$ and is
always less than 1. We are unable to estimate the Polyakov correlation
length, since the average of the Polyakov loop is always zero within
errors ($10^{-5}$), as expected if $\xi_P\sim \xi_z\lesssim 1$.
Again, data support the scenario that both gauge-invariant invariant
modes associated with $Q_{{\bm x},\mu}$ and gauge modes are massive in
the HT phase, even at the transition point, for any $N$, including
$N=\infty$, consistently with an $N=\infty$ first-order transition.

The results in LT phase are also in agreement with the scenario
reported in Sec.~\ref{sec2.2}.  The correlation length $\xi$ always
scales with $L$ for any $N$, as expected.  The correlation length
$\xi_z$ is instead finite in the infinite-volume limit and apparently
scales as $\xi_z \sim N$, a behavior which is also confirmed by the
Polyakov correlation length $\xi_P$.  Therefore, in the LT phase gauge
modes are massive for any finite $N$ and become massless in the limit
$N\to \infty$, consistently with what was observed on the LT side of
the transition.

\section{The large-$N$ standard solution: a critical discussion}
\label{largeN-sol}

The numerical data we have presented strongly suggest that the
CP$^{N-1}$ model with Hamiltonian (\ref{Ham}) undergoes a first-order
transition for any $N>2$ and also for $N=\infty$.  This is in contrast
with \cite{PV-19} that predicted a continuous transition using some
standard assumptions.  We will now review the large-$N$ calculations,
with the purpose of understanding which assumption is not correct.
For the specific Hamiltonian (\ref{Ham}), there is no need to use the
general approach of \cite{PV-19}. One can obtain the same results in a
more straighforward way, repeating on the lattice the same steps that
are used in continuum calculations (see \cite{MZ-03} and references
therein). This simply amounts to trivially extending to three
dimensions the lattice 2D calculations reviewed, for instance, in
\cite{CR-93}.

We start from the partition function, which can be written as
\begin{equation}
Z = \int \prod_{{\bm x}\mu} d\theta_{{\bm x}\mu} \,
   \prod_{\bm x} \left[d{\bm z}_{\bm x} d\bar{\bm z}_x 
    \delta( \bar{\bm z}_{\bm x} \cdot {\bm z}_{\bm x} - 1)\right] \, e^{-H},
\end{equation}
where we wrote $\lambda_{{\bm x}\mu} = \exp(i\theta_{{\bm x}\mu})$. As 
usual, we write 
\begin{equation}
\delta( \bar{\bm z}_{\bm x} \cdot {\bm z}_{\bm x} - 1) = 
   {\beta N \over 2 \pi i} \int_{c-i\infty}^{c+i \infty} d\gamma_x 
   \exp[-\beta \gamma_x N ( \bar{\bm z}_{\bm x} \cdot {\bm z}_{\bm x} - 1)].
\end{equation}
where $c$ is a real constant.
We can then integrate over the $\bm z$-fields, obtaining 
\begin{equation}
Z = (2\pi i)^{-V} \int \prod_{{\bm x}\mu} d\theta_{{\bm x}\mu} 
         \prod_{\bm x}  d\gamma_{\bm x} \, e^{-\beta N H_{\rm eff}},
\end{equation}
where 
\begin{equation}
H_{\rm eff}(\{\theta_{{\bm x}\mu},\gamma_{\bm x}\}) = 
   {1\over 2} \hbox{Tr}\ \log (B B^\dagger) + \sum_{\bm x} \gamma_{\bm x} ,
\end{equation}
the matrix $B_{\bm xy}$ is given by 
\begin{equation}
B_{\bm xy} = \gamma_{\bm x} \delta_{\bm xy} - 
    e^{i\theta_{\bm xy}} \eta_{\bm xy},
\end{equation}
and $\eta_{\bm xy} = 1$ is equal to 1 if $x$ and $y$ are nearest
neighbors and is zero otherwise.

The limiting behavior for $N\to \infty$ can be obtained by using the
usual saddle-point method. For this purpose we must determine the
stationary point of the effective Hamiltonian $H_{\rm eff}(\{
\theta_{{\bm x}\mu},\gamma_x \})$ with the lowest (free) energy.  In
the usual approach one {\em assumes} that the relevant stationary
point is obtained by considering {\em translation invariant} solutions
of the gap equations. In other words, the saddle point is obtained by
setting $\theta_{{\bm x}\mu} = \theta_0$ and $\gamma_x = \delta_0$,
where $\theta_0$ and $\delta_0$ are constants independent of the
position. Gauge invariance allows one to set $\theta_0 = 0$, implying
that the saddle point corresponds to setting $\lambda_{{\bm x}\mu} =
1$ on every link. Thus, the assumption of translation invariance
essentially implies that the gauge variables play no role in the large
$N$ limit (the same assumption is made in the continuum formulation,
see, e.g., \cite{MZ-03}). We thus obtain the effective Hamiltonian for
the large-$N$ O($2N$) vector theory.  One then predicts a continuous
transition located at
\begin{equation}
   \beta_{c,\infty} = \int {d^3p\over (2\pi)^3} {1\over \sum_\mu
\hat{p}_\mu^2} \approx 0.25273,
\label{betac-inf}
\end{equation}
where $\hat{p} = 2 \sin (p_\mu/2)$. In two dimensions, the assumption
turns out to be correct, see \cite{CR-93} for a review.  Our results
show instead that this is not the case in three dimensions. The
assumption that $\lambda_{{\bm x}\mu} = 1$ on every link for
$N=\infty$ is only correct in the LT phase. Indeed, in this phase we
observe $1/\xi_z, 1/\xi_P \sim 1/N$, which confirms the existence of a
massless gauge phase for $N=\infty$. In the HT phase, instead, even
for $N$ strictly equal to infinity, gauge fields are spatially
uncorrelated. This implies that, for small $\beta$, there is a
different non-translation invariant saddle point with a lower free
energy that gives the correct behavior of the theory.

As the HT phase is associated with a different saddle point of $H_{\rm
  eff}$, the critical point in the large-$N$ limit is not necessarily
given by Eq.~(\ref{betac-inf}). However, the presence of a single
transition, allows us to set the lower bound $\beta_c \ge
\beta_{c,\infty}$.  Indeed, in the opposite case, as the
translation-invariant saddle point gives the solution for all values
$\beta > \beta_c$, we would have a continuous transition for $\beta =
\beta_{c,\infty}$, with a finite correlation length $\xi$ in the
interval $\beta_c < \beta < \beta_{c,\infty}$.  As there is no
evidence of this intermediate phase, we conclude that $\beta_c \ge
\beta_{c,\infty}$.

\section{Conclusions} \label{Conclusions}

In this paper we have analyzed the phase diagram of the CP$^{N-1}$
model with Hamiltonian (\ref{Ham}), with the objective of
understanding the nature of the finite-temperature transition as a
function of the number $N$ of components. The numerical data indicate
that the transition is of first order. For all values of $N$ we
consider, $N>2$ up to $N=100$,
the correlation length $\xi$ obtained from correlations of
the gauge-invariant order parameter $Q^{ab}$ defined in
Eq.~(\ref{Q-def}) is of order one on the HT side of the transition and
diverges in the infinite-volume limit on the LT side.  Moreover, the
transition becomes stronger as $N$ increases: both the latent heat and
the surface tension, which parametrize the free energy barrier between
the two phases, increase with $N$.  Vector and gauge correlations are
massive for any finite $N$.  On the HT side of the transition, the
corresponding correlation lengths $\xi_z$ and $\xi_P$ are always of
order one, for any value of $N$: gauge fluctuations are always
uncorrelated, even for $N=\infty$.  In the low-temperature phase,
instead, we find that $\xi_z,\xi_P\sim N$, so that gauge modes become
massless in the large-$N$ limit.

Our results are consistent with a simple scenario in which, for any
$N$ including $N=\infty$, the HT phase is always disordered, up to the
transition point, where both correlations of the order parameter $Q$
and gauge correlations decay with a typical length scale of the order
of one lattice spacing. In the LT phase, $Q$ condenses, while
$\xi_z,\xi_P\sim N$, so that for $N=\infty$ both gauge-invariant and
gauge-dependent degrees of freedom are massless. The transition is
therefore of first order, even for $N=\infty$.

These results contradict the analytic predictions of the many papers
that investigated the large-$N$ limit, see, \cite{KS-08,PV-19} and
references therein. The disagreement can be traced back to one of the
standard assumptions which is used in the large-$N$ analysis, both for
continuum and lattice models \cite{CR-93,MZ-03,KS-08,PV-19}.  In the
calculation, one usually assumes that the relevant saddle point that
controls the behavior of the large-$N$ free energy is translation
invariant. For the gauge fields, this assumption implies that one can
set $\lambda_{{\bm x},\mu} = 1$ on any lattice link: gauge fields are
assumed to play no role for $N=\infty$. Our results show that the
assumption is correct in the LT phase, but fails in the HT phase: even
for $N=\infty$ the gauge fields are disordered for any $\beta <
\beta_c$.  This is essentially consistent with the results of
\cite{MS-90}, that observed that hedgehog configurations forbid the
ordering of gauge fields in the HT phase, at least if one takes the
limit $N\to\infty$ before the limit $\xi\to\infty$.

The present results are in agreement with the predictions obtained in
the so-called LGW approach defined in terms of the order parameter
$Q^{ab}$, provided one assumes that the presence of a $\Phi^3$ term in
the LGW Hamiltonian implies the absence of continuous transitions. It
should be stressed that this assumption should not be taken for
granted as it relies on an extrapolation of mean-field results to
three dimensions.  Note that, although we find no evidence of a
large-$N$ critical transition, our results do not exclude it either,
as it is a priori possible that our model is outside the attraction
domain of this elusive fixed point.

It is interesting to compare our results with those of
\cite{Kaul-12,KS-12,BMK-13} for SU($N$) quantum antiferromagnets.
Reference~\cite{Kaul-12} studied a bilayer two-dimensional system and
found a behavior analogous to what we find here. The transition is of
first order for $N\ge 4$ and becomes stronger as $N$ increases from
$N=4$ to $N=6$. On the other hand, for a single-layer two-dimensional
system \cite{KS-12}, an apparently continuous transition was always
observed. The main difference between the two models is the
topological nature of the allowed configurations.  In the bilayer
system, monopoles are allowed, while in the single-layer case
monopoles are suppressed.  In the model we consider, monopoles are
allowed and we expect the transition to be characterized by their
binding/unbinding: the monopole density should be positive in the HT
phase and vanishing in the LT phase. Thus, on the basis of the results
of \cite{Kaul-12,KS-12} and consistently with the discussion of
\cite{MS-90}, one may blame monopoles for the absence of a continuous
transition.  As the suppression of monopoles corresponds to adding an
ordering interaction in the HT phase, it is conceivable---this would
be consistent with the results of \cite{KS-12,BMK-13}---that a
continuous transition can be observed in a model in which monopoles
are completely, or at least partially, suppressed. Clearly, additional
work is needed to identify the role that monopoles play in the
large-$N$ limit.

\end{document}